%% file: Pi0A00.tex
\documentclass[aps,prc,twocolumn,amsmath,amssymb,superscriptaddress]{revtex4-1}
\usepackage{graphicx}
\usepackage{hyperref}
\usepackage{color}
\newcommand{\be}{\begin{equation}}
\newcommand{\ee}{\end{equation}}
\newcommand{\beq}{\begin{eqnarray}}
\newcommand{\eeq}{\end{eqnarray}}
\newcommand{\ba}{\begin{array}}
\newcommand{\ea}{\end{array}}

\begin{document}

\title{Pseudoscalar and Scalar Meson Photoproduction Interpreted by Regge Phenomenology}

\date{\today}

%--------------------- AUTHOR LIST ---------------------------
\input{author_list.tex}

%---------------------- ABSTRACT -----------------------------
\begin{abstract}
We have evaluated pseudoscalar and scalar neutral pion photoproduction in $\vec{\gamma}p\to\pi^0p$ and $\vec{\gamma}p\to a_0^0p$ above the resonance region and within Regge phenomenology.  Our fit, including GlueX $\Sigma$ pseudoscalar photoproduction data, shows that previous SLAC $\Sigma$ measurements for $\vec{\gamma}p \to \pi^0p$ above $E_\gamma = 4~\mathrm{GeV}$ are vary substantially from SLAC data with more recent measurements made by GlueX in vicinity of $E_\gamma = 9~\mathrm{GeV}$. The Regge model predicts that the beam polarization asymmetry $\Sigma$ of the scalar meson is opposite to that of pseudoscalar meson photoproduction, however, the cross sections are similar. While the natural parity vector meson exchange is dominant in both cases, the contribution of the unnatural parity pseudovector meson exchange is very small. Using Regge phenomenology, we predicted high energy behavior for double polarized observables $\mathbb{E}$, $\mathbb{F}$, $\mathbb{G}$, and $\mathbb{H}$ for the reactions $\gamma p\to \pi^0p$ and $\gamma p\to a_0^0p$.
\end{abstract}

\maketitle

%-------------------------Introduction-------------------------------------
The nature of light mesons is still a subject of controversy. The pseudoscalar, pseudovector (axial vector), vector, and scalar sectors of light-quark spectroscopy remain poorly understood. Various phenomenological models have been suggested to describe the light mesons and light meson photoproduction dynamics is actively studied  theoretically and experimentally.

The dynamics of high-energy light meson photoproduction has a long history (see, \textit{e.g.}, Refs.~\cite{Zweig:1964g, Goldstein:1973xn}). The recent development of a theoretical framework~\cite{Donnachie:2008nb, Laget:2010za, Yu:2011zu, Mathieu:2015eia, Donnachie:2015jaa, Kashevarov:2017vyl} was stimulated by  measurements at the Thomas Jefferson National Accelerator Facility (JLab), the ELectron Stretcher Accelerator (ELSA), and Super Photon Ring – $8~\mathrm{GeV}$ (SPring-8) for pseudoscalar~\cite{JeffersonLabHallA:2002vjd, Dugger:2007bt, CLAS:2009tyz, CLAS:2009wde, Chen:2009sda, CLAS:2016zjy, CLAS:2017dco, GlueX:2017zoo, CLAS:2017kyf, GlueX:2019adl, CLAS:2020vpr}, pseudovector~\cite{CLAS:2016zjy}, vector~\cite{CLAS:2001zxv, CLAS:2002cdi, CLAS:2009hpc, Dey:2014tfa, Ali:2019lzf}, and scalar~\cite{CB-ELSA:2008zkd, CLAS:2008ycy} meson photoproduction above the resonance region, where the Regge model is applicable to describe the dynamics of the light meson photoproduction.

The aim of this paper is to compare pseudoscalar neutral pion ($J^{PC} = 0^{-+}$) photoproduction in $\gamma p\to\pi^0p$ and scalar neutral $a_0(980)$ ($J^{PC} = 0^{++}$) photoproduction in $\gamma p\to a_0^0p$ in vicinity of $E_\gamma = 9~\mathrm{GeV}$.

Dynamically, a natural parity vector $\rho$ and $\omega$ meson exchange  holds the leading role in forward neutral pseudoscalar $\pi^0$ photoproduction, which leads, in particular, to +1 in $\Sigma$ beam asymmetry. Adding an unnatural parity pseudovector $h_1(1170)$ and $b_1(1235)$ meson exchange reduces the $\Sigma$ value. We note that very little is known about these pseudovector mesons~\cite{ParticleDataGroup:2022pth}. The $h_1(1170)$ and $b_1(1235)$ Regge trajectories are 1/2 unit below those of the $\rho$ and $\omega$, resulting in a $s^{1/2}$ difference between vector and pseudovector meson exchange. This means that contribution from $h_1(1170)$ and $b_1(1235)$ is suppressed with increasing photon energy.

From vector meson dominance~\cite{GellMann:1961tg, Sakurai:1969} $\omega$ exchange is expected
to be the dominant mechanism for $\pi^0$ photoproduction, the cross section for which is characterized by a pronounced dip at $|t|\sim 0.5~\mathrm{GeV^2}$ (see, \textit{e.g.}, Ref.~\cite{CLAS:2009hpc}). In a pure Regge model, this is usually attributed to a Nonsense Wrong-Signature Zero (NWSZ) for the $\omega$ residue~\cite{Chiu:1971wd}, where the Regge trajectory crosses 0 at negative {\it t}.

Neutral scalar $a_0^0$ photoproduction has 4 independent helicity amplitudes, similar to that of the  $\pi^0$~\cite{Barker:1975bp}.  These helicities satisfy the same kind of parity relations up to the \textit{sign} for leading order ($|t|$ / $s$) approximation. There are positive and negative combinations of helicities that have definite parity in the  $t$-channel. Thus, Regge model will yield a simple +1 for $\pi^0$ or --1 for $a_0^0$ predictions for $\Sigma$ beam asymmetry~\cite{Goldstein:1973xn, Worden:1972dc} (see Appendix for details). Overall, d$\sigma$/dt (unpolarized), $\mathbb{P}$, $\mathbb{E}$, and $\mathbb{F}$ are unchanged signs while $\Sigma$, $\mathbb{G}$, and $\mathbb{H}$ change signs.

\begin{figure*}[h]
\centering
\includegraphics[width=0.85\textwidth,keepaspectratio]{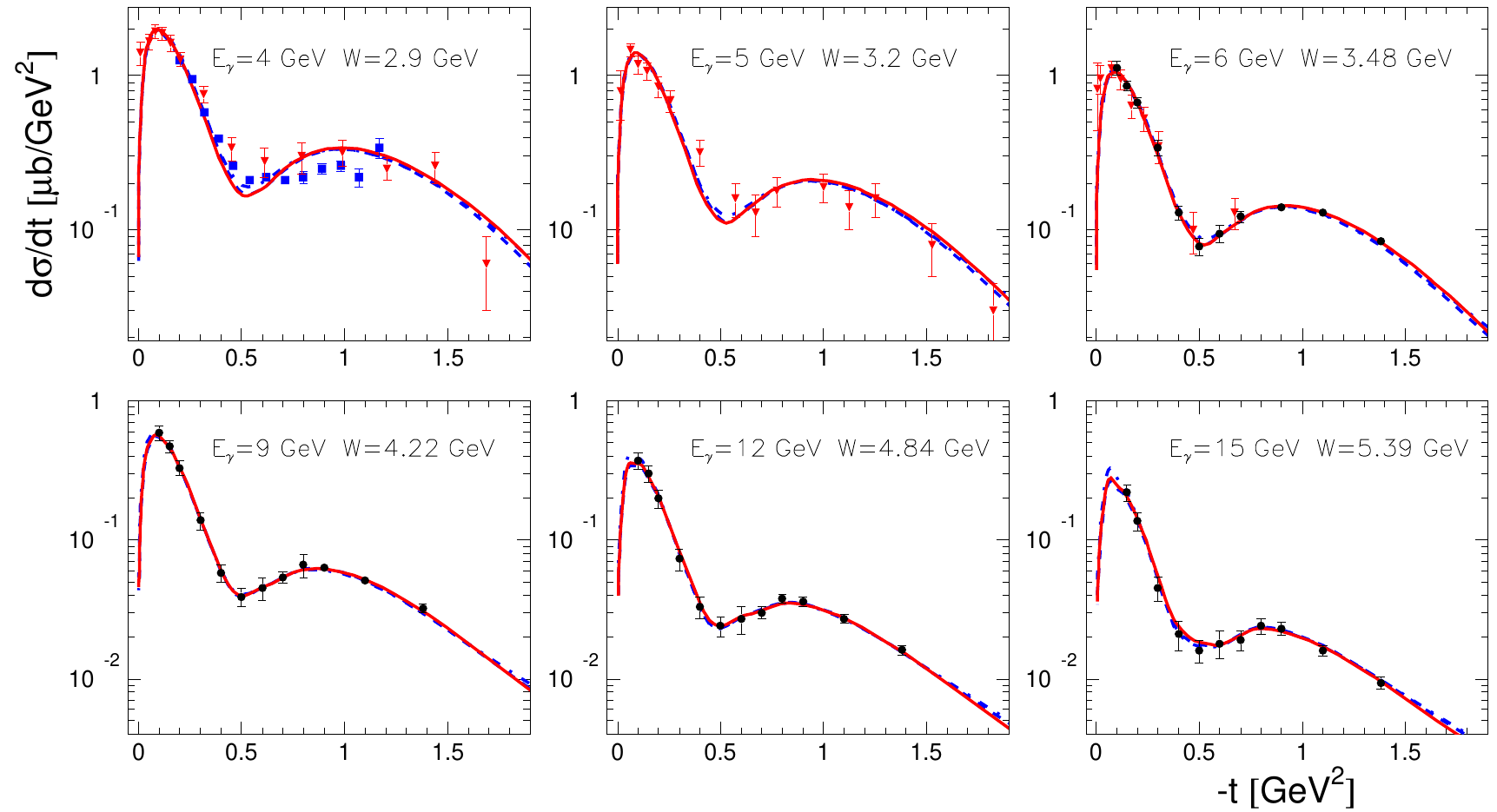}
\caption{Differential cross sections for $\gamma p \to \pi^0p$~\cite{Kashevarov:2017vyl}.  Data are from SLAC~\cite{Anderson:1971xh} (black filled circles), DESY~\cite{Braunschweig:1968pra} (red filled triangles), and \cite{Braunschweig:1973vu} (blue filled squares). Samples of Regge calculations above $4~\mathrm{GeV}$, using model~\cite{Kashevarov:2017vyl}, associated with item~\#1 (no GlueX $\Sigma$s, red solid curves), item~\#2 (GlueX $\Sigma$s and no other asymmetry data, blue dash-dotted curves), and item~\#3 (no SLAC $\Sigma$s, blue dashed curves) (see details in the text).}
\label{fig:11}
\end{figure*}

At high energies, the $|t|$ dependency of the $\pi^0$ differential cross section demonstrates a ``diffraction'' behavior. In particular, the observed dip at $|t|\sim 0.5~\mathrm{GeV^2}$ is visible in several independent measurements from SLAC, Deutsches Elektronen-SYnchrotron (DESY), and CLAS at $E_\gamma = 3$ to $15~\mathrm{GeV}$~\cite{Braunschweig:1968pra, Braunschweig:1973vu, Anderson:1971xh, CLAS:2017kyf} (Fig.~\ref{fig:11}). One can see that differing Regge approaches described cross sections quite well~\cite{Goldstein:1973xn, Laget:2010za, Yu:2011zu, Mathieu:2015eia, Donnachie:2015jaa, Mathieu:2015eia, Kashevarov:2017vyl}. The CBELSA differential cross section for $a_0^0$ photoproduction at energies above $E_\gamma = 2~\mathrm{GeV}$ shows similar $|t|$ distribution~\cite{CB-ELSA:2008zkd} (see, \textit{e.g.}, Fig.~3~(left) from Ref.~\cite{Donnachie:2015jaa}), as in the $\pi^0$ photoproduction as shown on Fig.~\ref{fig:11}.

\begin{figure*}[h]
\centering
\includegraphics[width=0.95\textwidth,keepaspectratio]{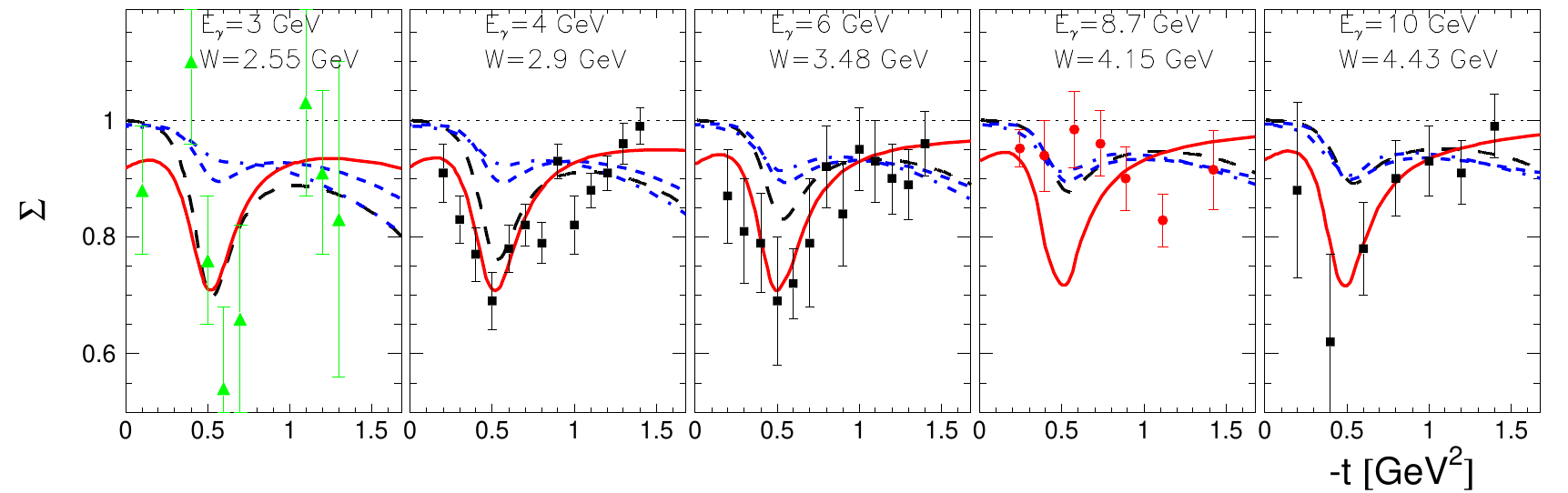}
\caption{Polarization observables $\Sigma$ for $\vec{\gamma} p\to \pi^0p$.  Data: SLAC~\cite{Anderson:1971xh} (black filled squares), GlueX~\cite{GlueX:2017zoo} (red filled circles), CEA~\cite{Bellenger:1969aa} (green filled triangles). Samples of Regge calculations, using model~\cite{Kashevarov:2017vyl}, associated with items~\#1 (no GlueX $\Sigma$s, red solid curves), item~\#2 (GlueX $\Sigma$s and no other asymmetry data, blue dash-dotted curves), item~\#3 (no SLAC $\Sigma$s, blue dashed curves) (see details in the text). Horizontal dotted lines correspond to $|\Sigma|$ = 1.}
\label{fig:12}
\end{figure*}

A similar dip at $|t|\sim 0.5~\mathrm{GeV^2}$ in $\Sigma$ beam asymmetry was reported by the Stanford Linear Accelerator Center (SLAC) at $E_\gamma = 4$, 6, and $10~\mathrm{GeV}$~\cite{Anderson:1971xh} (Fig.~\ref{fig:12}). Note that there is no correlation between dips in differential cross section and $\Sigma$ beam asymmetry. Here, $\Sigma$ is the linearly polarized photon asymmetry. Let us note that all previous theoretical calculations (before the JLab GlueX $\Sigma$ data for $\vec{\gamma}p\to \pi^0p$ came~\cite{GlueX:2017zoo}) were under influence of this dip at $|t|\sim 0.5~\mathrm{GeV^2}$ in $\Sigma$ beam asymmetries.

Recent GlueX $\Sigma$ measurements have shown that this dip is almost absent (or within uncertainties) at $E_\gamma = 8.7~\mathrm{GeV}$ (Fig.~\ref{fig:12}). An analysis of previous SLAC measurements~\cite{Anderson:1971xh} shows that the dip region in the SLAC data had serious background contributions from Compton scattering; it is plausible that these data might be unreliable in dip region.
%~\cite{Dugger:2022m}. 
Citation: \textit{``Since the experimental resolution was not sufficient to separate $\pi^0$ photoproduction from proton Compton scattering, it was necessary to make a further correction by subtracting the appropriate Compton contribution''}~\cite{Anderson:1971xh}, and the \textit{``extracted $\pi^0$ cross section in the dip depends very critically on this correction. For example, at $15~\mathrm{GeV}$ and $t = -0.5~\mathrm{(GeV/c)^2}$, two-thirds of the observed yield is due to Compton scattering''}~\cite{Anderson:1971xh}.

\begin{figure*}[h]
\centering
\includegraphics[width=0.90\textwidth,keepaspectratio]{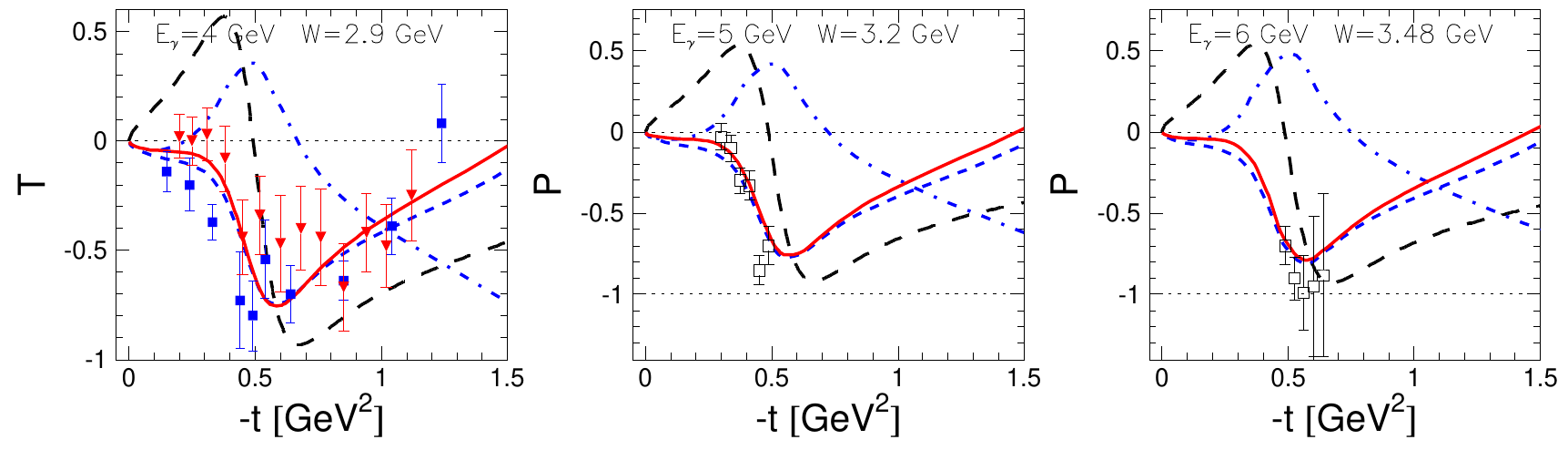}
\caption{Polarization observables $\mathbb{T}$  and $\mathbb{P}$ for $\gamma p\to \pi^0p$.  Data:  Daresbury~\cite{Booth:1972qp} (red filled triangles), DESY~\cite{Bienlein:1973pt} (blue filled squares), CEA~\cite{Deutsch:1972wyh} (black open squares). Samples of Regge calculations, using  model~\cite{Kashevarov:2017vyl}, associated with items~\#1 (no GlueX $\Sigma$s, red solid curves), item~\#2 (GlueX $\Sigma$s and no other asymmetry data, blue dash-dotted curves), item~\#3 (no SLAC $\Sigma$s, blue dashed curves), and Ref.~\cite{Yu:2011zu} (black long dashed curves)~(see details in the text). Horizontal dotted lines correspond to $|\mathbb{T}|$ = $|\mathbb{P}|$ = 0, 1.}
\label{fig:13}
\end{figure*}

%-------------------------------------Fitting data-------------------------------
To evaluate this $\Sigma$ case for $\vec{\gamma}p\to\pi^0p$ (\textit{i.e.}, why the energy sequence from $E_\gamma = 3$ to $10~\mathrm{GeV}$ is broken at $8.7~\mathrm{GeV}$), we performed three fits for available unpolarized and polarized data above $E_\gamma = 3~\mathrm{GeV}$, using a Regge model~\cite{Kashevarov:2017vyl}, to compare with (results shown on Figs.~\ref{fig:11} -- \ref{fig:13}): 
\begin{enumerate}
\item Differential cross sections from SLAC and DESY from $E_\gamma = 4$ to $15~\mathrm{GeV}$~\cite{Braunschweig:1968pra, Braunschweig:1973vu, Anderson:1971xh}, $\Sigma$ from SLAC at $E_\gamma = 4$, 6, and $10~\mathrm{GeV}$~\cite{Anderson:1971xh}, GlueX at $E_\gamma = 8.7~\mathrm{GeV}$~\cite{GlueX:2017zoo}, and Cambridge Electron Accelerator at Harvard/MIT (CEA) at $E_\gamma = 3~\mathrm{GeV}$~\cite{Bellenger:1969aa}; $\mathbb{T}$ from the electron synchrotron at the Daresbury Laboratory and DESY at $E_\gamma = 4~\mathrm{GeV}$~\cite{Booth:1972qp, Bienlein:1973pt}; and $\mathbb{P}$ from CEA at $E_\gamma = 5$ and $6~\mathrm{GeV}$~\cite{Deutsch:1972wyh} were fitted. Result looks good except the GlueX $\Sigma$ at $E_\gamma = 8.7~\mathrm{GeV}$ case.
\item Then, we have attempted to fit recent values of GlueX $\pi^0$ $\Sigma$~\cite{GlueX:2017zoo} in order to determine their effect on the Regge approach. We fitted the same differential cross sections. No SLAC $\Sigma$ and Daresbury with DESY $\mathbb{T}$ and CEA $\mathbb{P}$ polarized measurements were included in this fit. Obviously, the small weight of the GlueX vs. SLAC $\Sigma$ data (Fig.~\ref{fig:12}) does not allow one to feel comfortable with this fit because it does not predict the $\mathbb{T}$ with $\mathbb{P}$ polarized observables (see, \textit{e.g.}, Fig.~\ref{fig:13}).
\item Finally, we used  differential cross sections from SLAC and DESY; $\Sigma$ from GlueX; $\mathbb{T}$ from Daresbury and DESY; and $\mathbb{P}$ from CEA.  No SLAC $\Sigma$ polarized measurements were included in this fit. It results a good description all unpolarized and polarized measurements except predictions for SLAC $\Sigma$ (Figs.~\ref{fig:11} -- \ref{fig:13}).
\end{enumerate}

%----------------------------------------------------
%-----------------------------------------------------

%----------------------------------------------------
%----------------------------------------------------

%----------------------------------------------------
%----------------------------------------------------

One can see that differential cross sections for both pseudoscalar $\pi^0$ and scalar $a_0^0$ photoproduction look similar, while for the $\Sigma$ observable, there is no exact mirror reflection (opposite) because the dynamics of pseudoscalar $\pi^0$ and scalar $a_0^0$ photoproduction is different. The simple structure of $\Sigma$ for the latter reaction results from the production amplitude consisting of only the natural parity exchange, whereas the former reaction exhibits the nontrivial structure by the asymmetry between the natural and unnatural parity exchanges.

\begin{figure*}[h]
\centering
\includegraphics[width=0.85\textwidth,keepaspectratio]{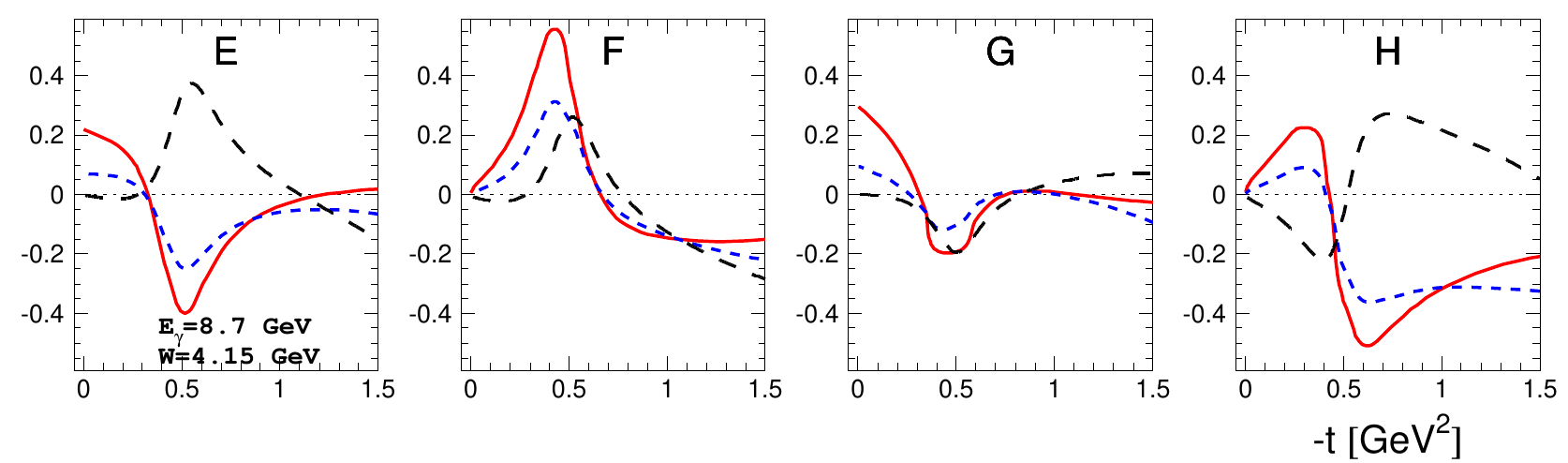}
\includegraphics[width=0.85\textwidth,keepaspectratio]{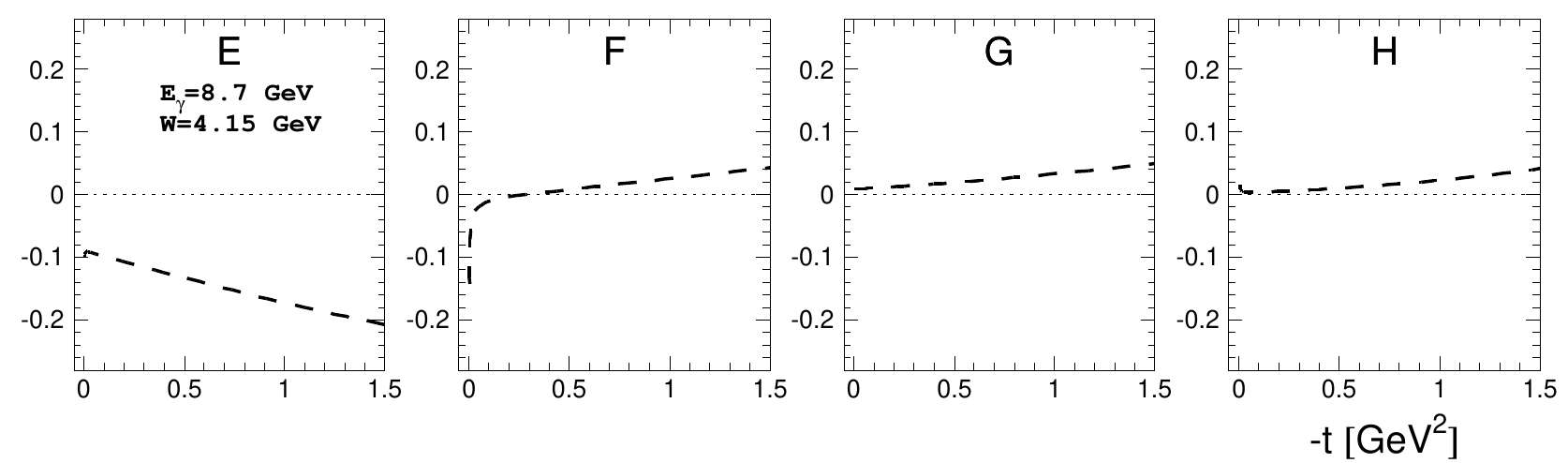}
\caption{Polarization observables $\mathbb{E}$, $\mathbb{F}$, $\mathbb{G}$, and $\mathbb{H}$ at $E_\gamma = 8.7~\mathrm{GeV}$ ($W = 4.15~\mathrm{GeV}$) for $\gamma p\to \pi^0p$ (top) and $\gamma p\to a_0^0p$ (bottom). Samples of Regge calculations using Regge model~\cite{Kashevarov:2017vyl}: associated with item~\#1 (no GlueX $\Sigma$s, red solid curves), item~\#3 (no SLAC $\Sigma$s, blue dashed curves), and predictions using model~\cite{Yu:2011zu} (black long dashed curves) (see details in the text). Horizontal dotted lines correspond to $|\mathbb{E}|$ = $|\mathbb{F}|$ = $|\mathbb{G}|$ = $|\mathbb{H}|$ = 0.}
\label{fig:14}
\end{figure*}

The use of polarized targets in conjunction with polarized beams allows a new set of measurements to be made which can often be instrumental in differentiating between various models. The target asymmetry, $\mathbb{T}$, is a measure of the azimuthal asymmetry when the target nucleon carries polarization in a direction perpendicular to the beam axis. $\mathbb{P}$, often termed the recoil polarization observable, is a measure of
the induced polarization of the recoiling nucleon (see Fig.~\ref{fig:14}).

Using circularly polarized photons one can measure the asymmetry with the proton target polarized along the beam ($\mathbb{E}$) or perpendicular to the beam in the reaction plane ($\mathbb{F}$). The linearly polarised photon beam and a longitudinally spin-polarized proton target allow access to $\mathbb{G}$, while $\mathbb{H}$ is determined from azimuthal asymmetries using measurements with linearly polarized photons and transversely polarized target nucleons. In the case of $\mathbb{G}$ and $\mathbb{H}$, photons polarized at $45^\circ$ to the reaction plane~\cite{Goldstein:1973xn} while now community does it for $90^\circ$. It results a flip the sign of quantities vs Ref.~\cite{Goldstein:1973xn}. Figure~\ref{fig:14} gives our predictions for 4 of these polarized quantities.

%----------------------------------------------------
%----------------------------------------------------

The set of predictions shown is very model dependent with a dominant role being played by the pole-cut phase relationship. As such, these measurements may provide sensitive tests for different cut models. Without making any detailed comparisons, we can note the following feature. Any NWSZ type model will have a peak in each curve at or near $|t|\sim 0.5~\mathrm{GeV^2}$ while a strong-cut model will give a zero for each measurement near $|t|\sim 0.5~\mathrm{GeV^2}$ (Fig.~\ref{fig:14}).

Polarization observables $\mathbb{E}$, $\mathbb{F}$, $\mathbb{G}$, and $\mathbb{H}$ for $\gamma p\to \pi^0p$ ($\gamma p\to a_0^0p$) at $8.7~\mathrm{GeV}$ shown on Fig.~\ref{fig:14} (top). One can see a reasonable agreement between predictions for polarized observables at $8.7~\mathrm{GeV}$. Fig.~\ref{fig:14} (bottom) shows predictions for polarization observables $\mathbb{E}$, $\mathbb{F}$, $\mathbb{G}$, and $\mathbb{H}$ for $\gamma p\to a_0^0p$ at $8.7~\mathrm{GeV}$ as well.

%----------------------------------------------------
\begin{figure}[h]
\centering
\includegraphics[width=0.47\textwidth,angle=0,keepaspectratio]{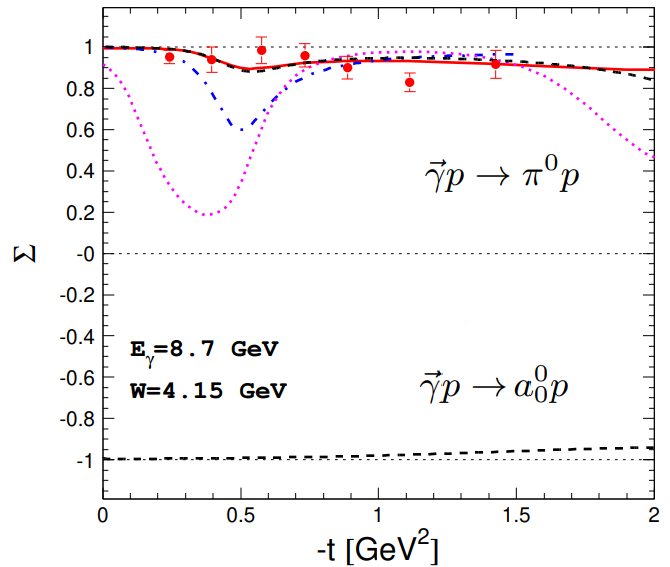}
\caption{Polarization observable $\Sigma$ for $\vec{\gamma} p\to \pi^0p$ and $\vec{\gamma} p\to a_0^0p$ at at $E_\gamma = 8.7~\mathrm{GeV}$ ($W = 4.15~\mathrm{GeV}$). GlueX $\pi^0$ data are from Ref.~\cite{GlueX:2017zoo}. Samples of Regge calculations are from Ref.~\cite{Goldstein:1973xn} (magenta dotted curve), Ref.~\cite{Donnachie:2015jaa} (blue dash-dotted curve), using model~\cite{{Kashevarov:2017vyl}} item~\#3 (no SLAC $\Sigma$s, red solid curve), and Ref.~\cite{Yu:2011zu} (black long dashed curves). Horizontal dotted lines correspond to $|\Sigma|$ = 0, 1.}
\label{fig:15}
\end{figure}
%----------------------------------------------------

%-------------------------------------Conclusion---------------------------------------
\underline{Finally}, we compare the beam polarization asymmetry  for $\pi^0$ production with that for the $a_0^0$ channel. Our third fit (see above) shows that previous SLAC $\Sigma$ measurements for $\vec{\gamma}p\to\pi^0p$ are not reliable and $\pi^0$ $\Sigma$ data is still under-constrained. It would be interesting to observe if the beam polarization asymmetry for the scalar meson is opposite to the pseudoscalar meson photoproduction as predicted  (Fig.~\ref{fig:15}). Let us note that the analysis of the GlueX $\Sigma$ data for $a_0^0$ photoproduction is in progress~\cite{Fegan:2020pyl}. It is crucial to get information about the radiative decay width of the scalar meson to vector and axial vector mesons and, hence, to be able to determine their coupling constants in order to have a reliable description of the dynamics of the reactions.  Obviously, to determine dynamics of scalar meson photoproduction, more cross sections for $a_0^0$ measurements are in order.

%-------------- Appendix -------------------------------------
\vspace{5mm}
\section{Appendix}

Comparing $\pi^0$ to $a_0^0$ photoproduction, there will be 4 independent helicity amplitudes (see Eq.~(5) from Ref.~\cite{Goldstein:1973xn}).

Parity relations for $2 \to 2$ helicity amplitudes in center-of-mass (CM) and X-Z plane (variables photon energy in laboratory frame $E_\gamma$ and meson polar production angle $\theta$ with meson azimuthal angle $\phi = \mathrm{0}$).

These come from applying the study by Jacob and Wick~\cite{Jacob:1959at} to particular cases summed over total two body reaction (see also Refs.~\cite{Goldstein:1981tj, Goldstein:1974dp}).

Using the notation of Ref.~\cite{Goldstein:1973xn}:
\begin{equation}\label{eq:a1}
    f_{-a,-b;-c,-d}(E_\gamma, \theta, \mathrm{0}) = \eta_g (-1)^{\lambda-\mu}f_{a,b;c,d}(E_\gamma, \theta) \>,
\end{equation}
where
\begin{equation}\label{eq:a2}
    \eta_g \equiv \frac{\eta_a \eta_b}{\eta_c \eta_d} (-1)^{s_c+s_d-s_a-s_b} \>,
\end{equation}
and
\begin{equation}\label{eq:a3}
    \lambda = a - b,~~~~~\mu = c - d \>.
\end{equation}

For pseudoscalar $\pi^0$ photoproduction
\begin{equation}\label{eq:a4}
    \eta_g = \frac{(- +)}{(- +)} (-1)^{-1} = - \>.
\end{equation}
For scalar $a_0^0$ photoproduction
\begin{equation}\label{eq:a5}
    \eta_g = \frac{(- +)}{(+ +)} (-1)^{-1} = + \>.
\end{equation}

Then, 4 helicity amplitudes in either case upper sign for pseudoscalar, lower for scalar parity phase).

\begin{equation}\label{eq:a6}
    f_{1,+1/2;0,+1/2} = f_1 = \pm f_{-1,-1/2;0-1/2} \>,
\end{equation}
\begin{equation}\label{eq:a7}
    f_{1,+1/2;0,-1/2} = f_2 = \mp f_{-1,-1/2;0+1/2} \>,
\end{equation}
\begin{equation}\label{eq:a8}
    f_{1,-1/2;0,+1/2} = f_3 = \mp f_{-1,+1/2;0-1/2} \>,
\end{equation}
\begin{equation}\label{eq:a9}
    f_{1,-1/2;0,-1/2} = f_4 = \pm f_{-1,+1/2;0+1/2} \>.
\end{equation}

To obtain leading $t$-channel exchange parity, combine the baryon vertices as
\begin{equation}\label{eq:a10}
    <d|~|b> \pm (-1)^{d-b}<-d|~|-b> \>
\end{equation}
for parity (naturality) positive or negative, respectively.

Then, the combinations that have leading positive or negative $t$-channel parity are constructed in the same way for pseudoscalar and scalar meson production.
\begin{equation}\label{eq:a11}
    f_1^\pm = f_1 \pm f_4 \>,
\end{equation}
\begin{equation}\label{eq:a12}
    f_2^\pm = f_2 \mp f_3 \>.
\end{equation}

What about the boson vertex $\gamma\to\pi^0$ or $a_0^0$~? Here, there has to be definite $C$-parity. For $\pi^0$ that restricts the exchange $C$-parity to negative. Hence $\omega$ and $\rho$ for vector exchange ($J^{PC}: 1^{- -}$) and $b_1(1235)$ and $h_1(1170)$ for axial vector exchange ($J^{PC}: 1^{+ -}$). Both have negative $C$. For $a_0^0$, the $C$-parity is also minus, but the parity is opposite. That implies that the axial and polar vector exchanges are switched. Hence the polarized beam photon asymmetry, $\Sigma$, will behave oppositely for $a_0^0$ as for $\pi^0$ (see, \textit{e.g.}, Fig.~\ref{fig:15}).

The relations among the different helicity amplitudes depend on details of the coupling of $a_0^0$ to $\gamma$ and the exchanged bosons - the residue functions in the Regge pole contributions.

%--------------- Acknowledgments --------------------------
\vspace{5mm}
\section{Acknowledgments}

We thank Sandy Donnachie, Stuart Fegan, Yulia Kalashnikova, David Mack, and Michael Ostrick for useful remarks and continuous interest in the paper.
This work was supported in part by the U.~S. Department of Energy, Office of Science, Office of Nuclear Physics under Award No.~DE-SC0016583, by EU Horizon 2020 research and innovation program, STRONG-2020 project, under grant agreement No.~824093, and the National Research Foundation of Korea Grant No.~NRF-2022R1A2B5B01002307.

%---------------------- REFERENCES -------------------------

\input{references.tex}
%-----------------------------------------------
\end{document}

%% file: author_list.tex
\author{\mbox{Igor~I.~Strakovsky}}
\altaffiliation{Corresponding author: \texttt{igor@gwu.edu}}
\affiliation{Institute for Nuclear Studies, Department of Physics, The 
	George Washington University, Washington, DC 20052, USA}

\author{\mbox{William~J.~Briscoe}}
\affiliation{Institute for Nuclear Studies, Department of Physics, The 
	George Washington University, Washington, DC 20052, USA}

\author{\mbox{Olga~Cortes~Becerra}}
\affiliation{Institute for Nuclear Studies, Department of Physics, The 
	George Washington University, Washington, DC 20052, USA}

\author{\mbox{Michael Dugger}}
\affiliation{Department of Physics and Astronomy, Arizona State University, 
    Tempe, AZ 85287, USA}

\author{\mbox{Gary~Goldstein}}
\affiliation{Department of Physics and Astronomy, 
    Tufts University, Medford, MA 02155, USA}

\author{\mbox{Victor~L.~Kashevarov}}
\affiliation{Institut f\"ur Kernphysik, University of Mainz, D-55099 Mainz, Germany}

\author{\mbox{Axel~Schmidt}}
\affiliation{Institute for Nuclear Studies, Department of Physics, The
        George Washington University, Washington, DC 20052, USA}

\author{\mbox{Peter Solazzo}}
\affiliation{Institute for Nuclear Studies, Department of Physics, The
        George Washington University, Washington, DC 20052, USA}

\author{\mbox{Byung-Geel Yu}}
\affiliation{Research Institute of Basic Sciences, Korea Aerospace University, 
    Goyang 10540, South Korea}

\noaffiliation

%% file: Pi0A00.bbl
\begin{thebibliography}{99}
%-----------------Theory-------------------------
\bibitem{Zweig:1964g}                                   % 1
    G.~Zweig,
    ``The reaction $\gamma N\to \pi N$ at high energies,''
    Nuov.\ Cim.\ \textbf{32}, 2113 (1964).
\bibitem{Goldstein:1973xn}                              % 2
    G.~R.~Goldstein and J.~F.~Owens,
    ``$\pi^0$ photoproduction in a weak-Regge-cut model,''
    Phys.\ Rev.\ D\ \textbf{7}, 865 (1973).
\bibitem{Donnachie:2008nb}                              % 3
    A.~Donnachie and Y.~S.~Kalashnikova,
    ``Scalar meson photoproduction,''
    Phys.\ Rev.\ C\ \textbf{78}, 064603 (2008).
\bibitem{Laget:2010za}                                  % 4
    J.~M.~Laget,
    ``Unitarity constraints on neutral pion electroproduction,''
    Phys.\ Lett.\ B\ \textbf{695} (2011), 199 (2011).
\bibitem{Yu:2011zu}                                     % 5
    B.~G.~Yu, T.~K.~Choi, and W.~Kim,
    ``Regge phenomenology of pion photoproduction off the nucleon at forward angles,''
    Phys.\ Rev.\ C\ \textbf{83}, 025208 (2011).
\bibitem{Mathieu:2015eia}                               % 6
    V.~Mathieu, G.~Fox, and A.~P.~Szczepaniak,
    ``Neutral pion photoproduction in a Regge model,''
    Phys.\ Rev.\ D\ \textbf{92}, 074013 (2015);
%\bibitem{JPAC:2016lnm}
    J.~Nys \textit{et al.} [JPAC Collaboration],
    ``Finite-energy sum rules in eta photoproduction off a nucleon,''
    Phys.\ Rev.\ D\ \textbf{95}, 034014 (2017).
\bibitem{Donnachie:2015jaa}                             % 7
    A.~Donnachie and Y.~S.~Kalashnikova,
    ``Photoproduction of $a_0(980)$ and $f_0(980)$,''
    Phys.\ Rev.\ C\ \textbf{93}, 025203 (2016).
\bibitem{Kashevarov:2017vyl}                            % 8
    V.~L.~Kashevarov, M.~Ostrick, and L.~Tiator,
    ``Regge phenomenology in $\pi^0$ and $\eta$ photoproduction,''
    Phys.\ Rev.\ C\ \textbf{96}, 035207 (2017).
%-----------------pseudoscalar--------------------------------------
\bibitem{JeffersonLabHallA:2002vjd}                     % 9
    L.~Y.~Zhu \textit{et al.} [Jefferson Lab Hall~A Collaboration],
    ``Cross-section measurement of charged pion photoproduction from hydrogen and deuterium,''
    Phys.\ Rev.\ Lett.\ \textbf{91}, 022003 (2003).
\bibitem{Dugger:2007bt}                                 % 10
    M.~Dugger, B.~G.~Ritchie, J.~P.~Ball, P.~Collins, E.~Pasyuk, R.~A.~Arndt, W.~J.~Briscoe, I.~I.~Strakovsky, R.~L.~Workman, G.~S.~Adams \textit{et al.}
    ``$\pi^0$ photoproduction on the proton for photon energies from 0.675 to 2.875-GeV,''
    Phys.\ Rev.\ C\ \textbf{76}, 025211 (2007).
\bibitem{CLAS:2009tyz}                                  % 11
    M.~Dugger \textit{et al.} [CLAS Collaboration],
    ``$\pi^+$ photoproduction on the proton for photon energies from 0.725 to 2.875-GeV,''
    Phys.\ Rev.\ C\ \textbf{79}, 065206 (2009).
\bibitem{CLAS:2009wde}                                  % 12
    M.~Williams \textit{et al.} [CLAS Collaboration],
    ``Differential cross sections for the reactions $\gamma p \to p \eta$ and $\gamma p \to p \eta'$,''
    Phys.\ Rev.\ C\ \textbf{80}, 045213 (2009).
\bibitem{Chen:2009sda}                                  % 13
    W.~Chen, T.~Mibe, D.~Dutta, H.~Gao, J.~M.~Laget, M.~Mirazita, P.~Rossi, S.~Stepanyan, I.~I.~Strakovsky, M.~J.~Amaryan \textit{et al.}
    ``A measurement of the differential cross section for the reaction $\gamma n \to \pi^- p$ from deuterium,''
    Phys.\ Rev.\ Lett.\ \textbf{103} (2009), 012301 (2009).
\bibitem{CLAS:2016zjy}                                  % 14
    R.~Dickson \textit{et al.} [CLAS Collaboration],
    ``Photoproduction of the $f_1(1285)$ meson,''
    Phys.\ Rev.\ C\ \textbf{93}, 065202 (2016).
\bibitem{CLAS:2017dco}                                  % 15
    P.~T.~Mattione \textit{et al.} [CLAS Collaboration],
    ``Differential cross section measurements for $\gamma n\rightarrow{\pi}^{-}p$ above the first nucleon resonance region,''
    Phys.\ Rev.\ C\ \textbf{96}, 035204 (2017).
\bibitem{GlueX:2017zoo}                                 % 16
    H.~Al~Ghoul \textit{et al.} [GlueX Collaboration],
    ``Measurement of the beam asymmetry $\Sigma$ for $\pi^0$ and $\eta$ photoproduction on the proton at $E_\gamma = 9$~GeV,''
    Phys.\ Rev.\ C \textbf{95}, 042201 (2017).
\bibitem{CLAS:2017kyf}                                  % 17
    M.~C.~Kunkel \textit{et al.} [CLAS Collaboration],
    ``Exclusive photoproduction of $\pi^0$ up to large values of Mandelstam variables $s$, $t$, and $u$ with CLAS,''
    Phys.\ Rev.\ C\ \textbf{98}, 015207 (2018).
\bibitem{GlueX:2019adl}                                 % 18
    S.~Adhikari \textit{et al.} [GlueX Collaboration],
    ``Beam asymmetry $\mathbf{\Sigma}$ for the photoproduction of $\mathbf{\eta}$ and $\mathbf{\eta^{\prime}}$ mesons at $E_\gamma$ = 8.8~GeV,''
    Phys.\ Rev.\ C\ \textbf{100}, 052201 (2019).
\bibitem{CLAS:2020vpr}                                  % 19
    T.~Hu \textit{et al.} [CLAS Collaboration],
    ``Photoproduction of $\eta$ mesons off the proton for $1.2 < E_\gamma < 4.7$~GeV using CLAS at Jefferson Laboratory,''
    Phys.\ Rev.\ C\ \textbf{102}, 065203 (2020).
%------------------------vector--------------------------------------
\bibitem{CLAS:2001zxv}                                  % 20
    M.~Battaglieri \textit{et al.} [CLAS Collaboration],
    ``Photoproduction of the $\rho^0$ meson on the proton at large momentum transfer,''
    Phys.\ Rev.\ Lett.\ \textbf{87}, 172002 (2001).
\bibitem{CLAS:2002cdi}                                  % 21
    M.~Battaglieri \textit{et al.} [CLAS Collaboration],
    ``Photoproduction of the $\omega$ meson on the proton at large momentum transfer,''
    Phys.\ Rev.\ Lett.\ \textbf{90}, 022002 (2003).
\bibitem{CLAS:2009hpc}                                  % 22
    M.~Williams \textit{et al.} [CLAS Collaboration],
    ``Differential cross sections and spin density matrix elements for the reaction $\gamma p \to p \omega$,''
    Phys.\ Rev.\ C\ \textbf{80}, 065208 (2009).
\bibitem{Dey:2014tfa}                                   % 23
    B.~Dey \textit{et al.} [CLAS Collaboration],
    ``Data analysis techniques, differential cross sections, and spin density matrix elements for the reaction $\gamma p \rightarrow \phi p$,''
    Phys.\ Rev.\ C\ \textbf{89}, 055208 (2014).
\bibitem{Ali:2019lzf}                                   % 24
  A.~Ali \textit{et al.} [GlueX Collaboration],
  ``First measurement of near-threshold $J/\psi$ exclusive photoproduction off the proton,''
  Phys.\ Rev.\ Lett.\ \textbf{123}, 072001 (2019).
%------------------------scalar--------------------------------------
\bibitem{CB-ELSA:2008zkd}                               % 25
    I.~Horn \textit{et al.} [CB-ELSA Collaboration],
    ``Study of the reaction $\gamma p \to p \pi^0 \eta$,''
    Eur.\ Phys.\ J.\ A\ \textbf{38}, 173 (2008).
\bibitem{CLAS:2008ycy}                                  % 26
    M.~Battaglieri \textit{et al.} [CLAS Collaboration],
    ``First measurement of direct $f_0(980)$ photoproduction on the proton,''
    Phys.\ Rev.\ Lett.\ \textbf{102}, 102001 (2009).
%--------------------------------------------------------------------
\bibitem{ParticleDataGroup:2022pth}                     % 27
    R.~L.~Workman \textit{et al.} [Particle Data Group],
    ``Review of Particle Physics,''
    Prog.\ Theor.\ Exp.\ Phys.\ \textbf{2022}, 083C01 (2022).
\bibitem{GellMann:1961tg}                               % 28
  M.~Gell-Mann and F.~Zachariasen,
  ``Form-factors and vector mesons,''
  Phys.\ Rev.\  {\bf 124}, 953 (1961).
\bibitem{Sakurai:1969}                                  % 29
  J.~J.~Sakurai, ``Currents and Mesons,'' (The University of Chicago Press, Chicago, 1969).
\bibitem{Chiu:1971wd}                                   % 30
    C.~B.~Chiu,
    ``The role of Regge poles and cuts in high-energy phenomenology,''
    Nucl.\ Phys.\ B\ \textbf{30}, 477 (1971).
\bibitem{Barker:1975bp}                                 % 31
    I.~S.~Barker, A.~Donnachie, and J.~K.~Storrow,
    ``Complete experiments in pseudoscalar photoproduction,''
    Nucl.\ Phys.\ B\ \textbf{95}, 347 (1975).
\bibitem{Worden:1972dc}                                 % 32
    R.~Worden,
    ``Regge models of forward pion and eta photoproduction,''
   Nucl.\ Phys.\ B\ \textbf{37}, 253 (1972).
%-------------------------------old data-------------------------------
\bibitem{Anderson:1971xh}                               % 33
    R.~L.~Anderson, D.~Gustavson, J.~R.~Johnson, I.~Overman, D.~Ritson, B.~H.~Wiik, and D.~Worcester,
    ``High-energy $\pi^0$ photoproduction from hydrogen with unpolarized and linearly polarized photons,''
    Phys.\ Rev.\ D\ \textbf{4}, 1937 (1971).
\bibitem{Braunschweig:1968pra}                          % 34
    M.~Braunschweig, W.~Braunschweig, D.~Husmann, K.~L\"ubelsmeyer, and D.~Schmitz,
    ``Single photoproduction of neutral $\pi$-mesons on hydrogen at small angles between 4 and 5.8~GeV,''
    Phys.\ Lett.\ B\ \textbf{26}, 405 (1968).
\bibitem{Braunschweig:1973vu}                           % 35
    W.~Braunschweig, W.~Erlewein, H.~Frese, K.~Luebelsmeyer, H.~Meyer-Wachsmuth, D.~Schmitz, and A.~Schultz Von Dratzig,
    ``Single photoproduction of neutral $\pi$ mesons on hydrogen in the forward direction at 4~GeV,''
    Nucl.\ Phys.\ B\ \textbf{51}, 167 (1973).
\bibitem{Bellenger:1969aa}                              % 36
    D.~Bellenger, R.~Bordelon, K.~Cohen, S.~B.~Deutsch, W.~Lobar, D.~Luckey, L.~S.~Osborne, E.~Pothier, and R.~Schwitters,
    ``Photoproduction of $\pi^0$ with plane polarized 3-GeV photons,''
    Phys.\ Rev.\ Lett.\ \textbf{23}, 540 (1969).
\bibitem{Booth:1972qp}                                  % 37
    P.~S.~L.~Booth, G.~R.~Court, B.~Craven, R.~Gamet, P.~J.~Hayman, J.~R.~Holt, A.~P.~Hufton, J.~N.~Jackson, J.~H.~Norem, and W.~H.~Range,
    ``Polarized target asymmetry for $\pi^0$ photoproduction at 4~GeV,''
    Phys.\ Lett.\ B\ \textbf{38}, 339 (1972).
\bibitem{Bienlein:1973pt}                               % 38
    H.~Bienlein, W.~Braunschweig, H.~Dinter, W.~Erlewein, H.~Frese, J.~Knuetel, K.~Luebelsmeyer, S.~Mango, H.~Meyer-Wachsmuth, C.~C.~Morehouse \textit{et al.}
    ``Target asymmetry for neutral $\pi$ meson photoproduction on polarized protons at 4~GeV in the forward direction,''
    Phys.\ Lett.\ B\ \textbf{46} (1973), 131 (1973).
\bibitem{Deutsch:1972wyh}                               % 39
    M.~Deutsch, L.~Golub, P.~Kijewski, D.~Potter, D.~J.~Quinn, and J.~Rutherfoord,
    ``Recoil-proton polarization in neutral-pion photoproduction and in proton Compton scattering,''
    Phys.\ Rev.\ Lett.\ \textbf{29}, 1752 (1972).
%------------------------------------------------------------------
\bibitem{Fegan:2020pyl}                                 % 40
    S.~Fegan [for GlueX Collaboration],
    ``Beam asymmetries from light scalar meson photoproduction on the proton at GlueX,''
    AIP\ Conf.\ Proc.\ \textbf{2249}, 030007 (2020).
%-------------------------------------------appendix---------------
\bibitem{Jacob:1959at}                                  % 41
    M.~Jacob and G.~C.~Wick,
    ``On the general theory of collisions for particles with spin,''
    Annals\ Phys.\ \textbf{7}, 404 (1959).
\bibitem{Goldstein:1981tj}                              % 42
    G.~R.~Goldstein and M.~J.~Moravcsik,
    ``Symmetry constraints in optimal polarization formalisms with an application to $p p$ scattering,''
    Annals\ Phys.\ \textbf{142}, 219 (1982).
\bibitem{Goldstein:1974dp}                              % 43
    G.~R.~Goldstein, J.~F.~Owens, J.~P.~Rutherfoord, and M.~J.~Moravcsik,
    ``Spin correlation measurements in pseudoscalar meson photoproduction,''
    Nucl.\ Phys.\ B\ \textbf{80}, 164 (1974).
%\bibitem{Dugger:2022m}
%    M.~Dugger, private communications, 2022.
%\bibitem{Yu:2022bg}
%    B.~G.~Yu, private communication, 2022.
\end{thebibliography}
